\documentclass[11pt]{article}

\usepackage[dvips]{graphicx}
\usepackage{amsmath}
\usepackage{amsthm}
\usepackage{citesort}
\usepackage{amssymb}

\newtheoremstyle{test}
     {}
     {}
     {\rm}
     {}
     {\bf}
     {}
     { }
     {}
\theoremstyle{test}

\newtheorem{thm}{Theorem}
\newtheorem{ex}[thm]{Example}
\newtheorem{de}[thm]{Definition}

\newtheorem{rem}[thm]{Remark}

\newcommand{\pd}{partially decryptable}

\newcommand{\AS}{\Gamma}
\newcommand{\as}{{\cal A}}

\renewcommand{\l}{\ell}
\newcommand{\QED}{\hfill $\Box$}

\newcommand{\halflineskip}{\vspace*{0.5 \baselineskip}}

\newcommand\sA{\mbox{\boldmath$A$}}
\newcommand\ssA{\hspace*{-.5mm}\mbox{\boldmath\scriptsize$A$}}
\newcommand\sB{\mbox{\boldmath$B$}}

\newcommand\sI{\mbox{\boldmath$I$}}

\newcommand\sS{\mbox{\boldmath$S$}}

\newcommand\sV{\mbox{\boldmath$V$}}
\newcommand\sR{\mbox{\boldmath$R$}}
\newcommand\VV{2^{\mbox{\boldmath\scriptsize$V$}}}
\newcommand\ssV{\mbox{\boldmath\scriptsize$V$}}

\newcommand{\DEF}{\stackrel{\rm def}{=}}

\makeatletter
\def\eqnarray{%
        \stepcounter{equation}%
        \let\@currentlabel=\theequation
        \global\@eqnswtrue\global\@eqcnt\z@
        \tabskip\@centering
        \let\\=\@eqncr
        $$\halign to \displaywidth\bgroup\@eqnsel\hskip\@centering
        $\displaystyle\tabskip\z@{##}$&\global\@eqcnt\@ne
        \hfil$\displaystyle{{}##{}}$\hfil
        &\global\@eqcnt\tw@$\displaystyle\tabskip\z@{##}$\hfil
        \tabskip\@centering&\llap{##}\tabskip\z@\cr}
\makeatother

\title{Strongly Secure Ramp Secret Sharing Schemes \\
for General Access Structures}

\author{
Mitsugu Iwamoto\thanks{
Graduate School of Information Systems, University of
Electro-Communications, 1-5-1 Chofugaoka, Chofu-shi, Tokyo, 182-8585,
Japan. E-mail: {\tt mitsugu@hn.is.uec.ac.jp }
}~~~and~ 
Hirosuke Yamamoto\thanks{
Graduate School of Frontier Science, 
 University of Tokyo. 5-1-5 Kashiwanoha, Kashiwa-shi, Chiba 277-8561,
 Japan.
}}
\begin{document}

\maketitle

\begin{abstract}
Ramp secret sharing (SS) schemes  can be classified into
strong ramp SS schemes and weak ramp SS schemes.
The strong ramp SS schemes do not leak out any part of a secret
explicitly even in the case where some information about
the secret leaks from a non-qualified set of shares, and hence, 
they are more desirable than weak ramp SS schemes. However,
it is not known how to construct the strong ramp SS schemes
in the case of general access structures.
In this paper, it is shown that a strong ramp SS scheme
can always be constructed from a SS scheme with plural secrets
for any feasible general access structure. As a byproduct, it is pointed
 out that threshold ramp SS schemes based on Shamir's polynomial
 interpolation method are {\em not} always strong.
\end{abstract}
\section{Introduction}
A secret sharing (SS) scheme \cite{S-cacm,B-afips} is a method to encode
a secret $\sS$ into $n$ shares each of which has no information of
$\sS$, but $\sS$ can be decrypted by collecting several shares. For
example, a $(k,n)$-threshold SS scheme means that any $k$ out of $n$
shares can decrypt secret $\sS$ although any $k-1$ or less shares do
not leak out any information of $\sS$. The $(k,n)$-threshold access
structure can be generalized to so-called {\em general access
structures} which consist of the families of {\em qualified sets} and {\em
forbidden sets}. A qualified set is the subset of shares that can decrypt
the secret, but any information does not leak out from any forbidden
set. Generally, the efficiency of SS schemes is evaluated by the entropy
of each share, and it must hold that $H(V_i)\ge H(\sS)$ where
$H(\sS)$ and $H(V_i)$ are the entropies of secret $\sS$ and shares $V_i$,
$i=1,2,\ldots,n$, respectively \cite{KGH-it,CSGV-jc}. 

In order to improve the efficiency of SS schemes, {\em ramp} SS schemes
are proposed, which have a trade-off between security and coding
efficiency
\cite{BM-crypto85,HY-ieice,KOSOT-ecrypt93,OK-jucs,OK-acrypt94}. For
instance, in the $(k,L,n)$-threshold ramp SS scheme
\cite{BM-crypto85,HY-ieice}, we can decrypt $\sS$ from arbitrary
$k$ or more shares, but no information of $\sS$ can be obtained from any
$k-L$ or less shares. Furthermore, we assume that arbitrary $k-\ell$
shares leak out about $\sS$ with equivocation $(\ell/L) H(\sS)$ for
$\ell=1,2,\ldots,L$. In the case where $L=1$, the $(k,L,n)$-threshold 
SS scheme reduces to the ordinal $(k,n)$-threshold ramp SS scheme. Hence, to
distinguish ordinal SS schemes with ramp SS schemes, we call ordinal
SS schemes {\em perfect} SS schemes.  For any $(k,L,n)$-threshold access
structure, we can realize that $H(V_i) = H(\sS)/L$ \cite{HY-ieice}, and
hence, ramp SS schemes are more efficient than perfect SS schemes
\cite{BM-crypto85,HY-ieice}. Furthermore, ramp schemes with general
access structures are studied in
\cite{KOSOT-ecrypt93,OK-acrypt94,OK-jucs}. 

Since non-forbidden sets with $1 \le \ell \le L-1$ in ramp SS schemes
are allowed to leak out a part of a secret, it is important to analyze
how the secret partially leaks out. For example, if a secret is a
personal data that consists of name, address, job, income, bank account, etc.,
any part of the secret should not leak out explicitly. However, in the
case that the security is measured by the conditional entropy, we
cannot know whether or not some part of the secret can be decrypted from a
non-forbidden set. Hence, Yamamoto introduced the notion of {\em strong}
and {\em weak} ramp SS schemes \cite{HY-ieice}. A ramp SS scheme is
called a strong ramp SS scheme if it
does not leak out any part of a secret explicitly from any arbitrarily 
$k-\ell$ shares for $\ell=1,2,\ldots,L$. A ramp SS scheme is weak if 
it is not strong. But, it is not given how to construct strong ramp
SS schemes for arbitrary given general access structures although it is
known for $(k,L,n)$-threshold ramp SS schemes in \cite{HY-ieice}.

In this paper, we discuss strong ramp SS schemes with general access
structures. In section 2, we define ramp SS schemes
called {\em partially decryptable} (PD) ramp SS schemes, in which every
non-qualified set with $k-\ell$ shares can decrypt explicitly
$(L-\ell)/L$ parts of a secret. Then, we clarify the relation between PD
ramp SS schemes and perfect SS schemes with plural secrets. We also
point out that $(k,L,n)$-ramp SS schemes based on Shamir's polynomial
interpolation method are not always strong. Next, in section
3, we propose how to convert PD ramp SS schemes into strong ramp SS
schemes by using a linear transformation, and we clarify that any 
access structure that can be realized as a weak ramp SS scheme can also be
realized as a strong ramp SS scheme.

\section{Background and Preliminaries}
Let $\sV =\{V_1,V_2,\ldots,V_n\}$ be the set of all shares, and let
$2^{\ssV}$ be the family of all the subsets of $\sV$. Denote a secret by
an $L$-tuple $\sS=\{S_1,S_2,\ldots,S_L\}$, and each element of $\sS$ is
assumed to be a mutually independent random variable according to the
uniform distribution which takes values in a finite field
$\mathbb{F}$. We assume that $|\mathbb{F}|$ is sufficiently 
large\footnotemark\footnotetext{Throughout this paper, a set of shares
and a family of share sets are represented by upper case bold-face and
calligraphic font letters, respectively. For simplicity of notation, we
use $\sA\sB$ to represent  $\sA\cup\sB$ for sets $\sA$ and $\sB$, and 
$\{V\}$ is represented as $V$.  For example, $\sA
V=\sA\cup\{V\}$. Furthermore, let $\sA-\sB$ be a difference set of $\sA$
and $\sB$, and the cardinality of a set $\sA$ is denoted by $|\sA|$.}. Then,
denote by $H(\sS)$ and $H(\sA)$ the entropies of the secret $\sS$ and a
set of shares $\sA \subseteq \sV$, respectively. 

For families $\as_\ell\subseteq \VV$, $\ell=0,1,\ldots,L$, which consist
of subsets of $\sV$, we define ramp SS schemes as 
follows:

\begin{de}\label{ramp.def}
Let $\sS$ and  $\AS_L=\{\as_0,\as_1,\ldots,\as_L\}$ be a given secret
 and a given access structure. Then, $\{\sS,\sV,\AS_L\}$ is called a
 {\em ramp secret sharing} (SS) scheme if every subset $\sA\in\as_\ell$
 satisfies the following for $\ell=0,1,\ldots,L$.
\begin{eqnarray}\label{1}
H(\sS|\sA)=\frac{L-\ell}{L}H(\sS).
\end{eqnarray}
\QED
\end{de}

Equation (\ref{1}) implies that secret $\sS$ leaks out from any set
$\sA\in\as_\ell$ with the amount of $(\ell/L)H(\sS)$. Especially, $\sS$
can be completely decrypted from any $\sA \in \as_L$, but any
$\sA\in\as_0$ leaks out no information of $\sS$. Hence, in the case of
$L=1$, ramp SS schemes reduce to perfect SS schemes. Without loss of
generality, we can assume that $\as_\ell\neq\as_{\ell'}$ holds for $\ell \neq
\ell'$. Furthermore, we also assume that
$\bigcup_{\ell=0}^L\as_\ell=2^{\ssV}$. 

For example, an access structure of a $(k,L,n)$-ramp SS scheme
\cite{BM-crypto85,HY-ieice} can be defined as $
\as_0=\{\sA : 0 \le |\sA| \le k-L \}$, $\as_\ell=\{\sA :
|\sA|=k-L+\ell\}$ for $1 \le \ell \le L-1$, and $\as_L=\{\sA : k \le
|\sA| \le n\}$. 
It is shown in \cite{KOSOT-ecrypt93} that ramp SS schemes with general
access structures can be constructed if and only if the following
conditions are satisfied.

\begin{thm}[\cite{KOSOT-ecrypt93}]\label{KOSOT.thm}
A ramp SS scheme with access structure
 $\AS_L=\{\as_0,\as_1,\ldots,\as_L\}$ can be
 constructed if and only if each $\tilde{\cal A}_\ell \stackrel{\rm def}{=}
\bigcup_{k=\ell}^L\as_k,\ell=1,2,\ldots,L$ satisfies the {\em
 monotonicity} in the following sense: 
\begin{eqnarray}\label{kurosawa.eq}
\sA\in\tilde{\cal A}_\ell~ \Rightarrow~ \sA' \in \tilde{\cal A}_\ell 
~\mbox{for all}~\sA' \supseteq \sA.
\end{eqnarray}
\QED
\end{thm}

In the case of $L=1$, (\ref{kurosawa.eq}) in Theorem \ref{KOSOT.thm}
coincides with the necessary and sufficient condition to realize a 
perfect SS scheme with an access structure $\AS_1=\{\as_0,\as_1\}$, which
is proved in \cite{ISN-globcom}бе

From Theorem \ref{KOSOT.thm}, the {\em minimal} access structure $\as^-_\ell$,
$\ell=1,2,\ldots,L$ can be defined as follows:
\begin{eqnarray}\label{condition}
\as^-_\ell=\{\sA\in\as_\ell:\sA-\{V\} \not\in\as_\ell~\mbox{for~any}~V\in\sA\}.
\end{eqnarray}
\noindent
{\it Proof of Theorem \ref{KOSOT.thm} (\cite{KOSOT-ecrypt93}):}
We will prove only the sufficiency of (\ref{kurosawa.eq}) because the
necessity is clear. Let
$\sS=\{S_1,S_2,\ldots,S_L\}$ be a secret. From \cite{ISN-globcom}, in the
case that (\ref{kurosawa.eq}) holds, we can construct a perfect SS
scheme for the secret $S_\ell$ with the access structure
$\tilde{\AS}_\ell\stackrel{\rm def}{=}
\{\VV-\tilde{\cal A}_\ell, \tilde{\cal A}_\ell\}$
for every $\ell=1,2,\ldots,L$. Then, let
$\tilde{\sV}_{\ell}\DEF\{V_{\ell,1},V_{\ell,2},\ldots,V_{\ell,n}\}$ be
the set of whole shares for such a perfect SS scheme with access structure
$\tilde{\AS}_\ell$ for the secret $S_\ell$. 

Now, we define  $\sV_i\DEF\{V_{1,i},V_{2,i},\ldots,V_{L,i}\}$ by
 collecting the $i$-th share of $\tilde{\sV}_\ell$, $\ell=1,2,\ldots,L$. Then,
 it is easy to check that the share set
 $\sV=\{\sV_1,\sV_2,\ldots,\sV_n\}$ realizes the ramp SS scheme with access
 structure $\AS_L$ for the secret $\sS$. 
In this case,
we can decrypt $\{S_1,S_2,\ldots,S_\ell\}$ from a share set
$\sA\in\tilde{\as}_\ell$, although $\sA$ cannot obtain any information
of $\{S_\ell,S_{\ell+1},\ldots,S_L\}$, and hence, (\ref{1}) is
satisfied. \QED

In ramp SS schemes, the {\em coding rate} of the $i$-th share can be
defined as $\rho_i \DEF H(V_i)/H(\sS)$. To realize efficient ramp SS
schemes, each coding rate of a ramp SS scheme should be as small as
possible. Furthermore, it is known that $\rho_i \ge 1/L$ must hold for each
$i=1,2,\ldots,n$ in any ramp SS scheme with $L$-level access structure
$\AS_L$ \cite{HY-ieice,KOSOT-ecrypt93}. From this viewpoint, the ramp SS
 schemes shown in the proof of Theorem \ref{KOSOT.thm} are not
 efficient. On the contrary, Okada-Kurosawa \cite{OK-acrypt94} presented the
following example of a ramp SS scheme with a general access structure,
which is more efficient than the ramp SS scheme shown in the proof of
Theorem \ref{KOSOT.thm}.

\begin{ex}[\cite{OK-acrypt94}]\label{Okada.ex}
Consider the following access structure $\AS^{\rm ex}_2$ for a set of
 shares $\sV= \{V_1,V_2,V_3,$\\ $V_4\}$. 
\begin{eqnarray}
\label{OkK1.eq1}
\as^-_1 &=& \{\{V_1,V_4\},\{V_2,V_4\}\}, \\
\label{OkK1.eq3}
\as^-_2 &=& \{\{V_1,V_2,V_3\}\}.
\end{eqnarray}
Then, by letting the secret be $\sS=\{S_1,S_2\}$, a ramp SS scheme for the
 access structure $\AS^{\rm ex}_2$ in (\ref{OkK1.eq1}) and
 (\ref{OkK1.eq3}) can be realized as 
\begin{eqnarray}
\label{ex.eq1}
V_1&=&\{R_1,R_3\},\\
V_2&=&\{R_2,R_4\},\\
V_3&=&\{R_1+R_4+S_1,R_2+R_3+S_2\},\\
\label{ex.eq3}
V_4&=&\{R_1+S_1,R_2+S_1\},
\end{eqnarray}
where $R_1,R_2$ and $R_3$ are mutually independent random numbers which
 take values in the same finite field  $\mathbb{F}$. \QED
\end{ex}

From Example \ref{Okada.ex}, it is clear that the secret $S_2$ can be
decrypted from $\{V_1,V_4\}$, but any information of $S_1$ cannot be
obtained from the set. Hence, since $S_1$ and $S_2$ are mutually
independent, it holds that $H(\sS|V_1V_4)=H(S_1)=H(\sS)/2$. 
In this way, if the partial information of the secret can be explicitly
decrypted from every non-qualified set of shares, it is easy to
calculate the amount of leaked information. Furthermore, we also note
that such a ramp SS scheme can be considered as a special case of
perfect SS schemes with $L$ plural secrets
\cite{BSV-stacs93,BSCGV-crypto94,C-tcs03}. 

In SS schemes with plural secrets, we assume that secret information \
is given by an $L$-tuple $\sS^{(L)}=\{S^{(1)},S^{(2)},\ldots,S^{(L)}\}$ where $S^{(\ell)}$ are mutually independent random
variables. 
Then, an access structure for the secret $\sS^{(L)}$ is given by 
$\AS^{(L)}\DEF\{\as^{(1)},\as^{(2)},\ldots,\as^{(L)}\}$ where the secret
$S^{(\ell)}$ can be decrypted from any set in $\as^{(\ell)}
\subseteq \VV$ for $\ell=1,2,\ldots,L$ while no information of
$S^{(\ell)}$ can be obtained from any set $\sA\not\in\as^{(\ell)}$.

The SS schemes for $L$ secrets with an access structure $\AS^{(L)}$
can be defined as follows:

\begin{de}[\cite{BSCGV-crypto94}]\hspace*{-1.5mm}\footnotemark\footnotetext{
In the definition of SS schemes with plural secrets in
 \cite{BSCGV-crypto94}, it is assumed that $S_\ell$,
 $\ell=1,2,\ldots,L$, are not always mutually independent. But, we can
 reduce the definition in \cite{BSCGV-crypto94} to Definition
 \ref{plural-SS-ori.def}, in which $S_\ell$'s are mutually independent.}
 \label{plural-SS-ori.def}
Let $\AS^{(L)}=\{\as^{(1)},\as^{(2)},\ldots,\as^{(L)}\}$ be an access
 structure for $L$ secrets denoted by 
 $\sS^{(L)}=\{S^{(1)},S^{(2)},\ldots,S^{(L)}\}$. Then,
 $\{\sS^{(L)},\sV,\AS^{(L)}\}$ is called a SS scheme with $L$ secrets if
 it satisfies for all $\l = 1,2,\ldots L$ that  
\begin{eqnarray}
H(S^{(\ell)}|\sA) &=& 0~~~~~~~~~~~\mbox{for any}~\sA\in\as^{(\l)},\\
H(S^{(\ell)}|\sA') &=& H(S^{(\ell)})~~\hspace*{.6mm}\mbox{for any}~\sA'\not\in\as^{(\ell)}.
\end{eqnarray}
\QED
\end{de}

From \cite{BSCGV-crypto94}, Definition \ref{plural-SS-ori.def} is
equivalent to the following definition.

\begin{de}[\cite{BSCGV-crypto94}] \label{plural-SS.def}
Let $\AS^{(L)}=\{\as^{(1)},\as^{(2)},\ldots,\as^{(L)}\}$ be an access
 structure for $L$ secrets denoted by 
 $\sS^{(L)}=\{S^{(1)},S^{(2)},\ldots,S^{(L)}\}$.   
Let $\sS^{({\ssA})}\subseteq\sS$ be a subset of the secret that can be
 decrypted from a share set $\sA\subseteq\sV$ according to $\AS^{(L)}$,
 and we define that $\overline{\sS^{({\ssA})}}\stackrel{\rm def}{=}
\sS-\sS^{({\ssA})}$. Then, $\{\sS^{(L)},\sV,\AS^{(L)}\}$ is called a
 SS scheme with plural secrets $\sS^{(L)}$ if it satisfies that
\begin{eqnarray}
\label{sss-plural.eq1}
H\left(\left.\sS^{({\ssA})}\right|\sA\right) &=& 0, \\
\label{sss-plural.eq2}
H\left(\left.\hspace{.3mm}
\overline{\sS^{({\ssA})}}\hspace{.3mm}\right|\sA\right) &=& 
H\left(\hspace{.3mm}\overline{\sS^{(\ssA)}}\hspace{.3mm}\right),
\end{eqnarray}
for all $\sA\subseteq\sV$. \QED
\end{de}

Based on Definition \ref{plural-SS.def}, we define the \pd{} ramp SS
schemes that characterize the ramp SS schemes shown in the proof of
Theorem \ref{KOSOT.thm} and Example \ref{Okada.ex}.

\begin{de}\label{partial-SS.def}
Let $\sS=\{S_1,S_2,\ldots,S_L\}$ be secrets for an access structure
 $\AS_L=\{\as_1,\as_2,\ldots,\as_L\}$. Then,  $\{\sS,\sV,\AS_L\}$ is
 called a {\em partially decryptable} (PD) ramp SS scheme if there exists a
 part of the secret information $\sS_{\ssA}\subseteq\sS$ satisfying that 
\begin{eqnarray}
\label{number_of_secrets}
|\sS_{\ssA}|&=&\ell\\
\label{weak-ramp1.eq}
H(\sS_{\ssA}|\sA) &=& 0, \\
\label{weak-ramp2.eq}
H\left(\hspace{.3mm}\overline{\sS_{\ssA}}|\sA\hspace{.3mm}\right) &=& 
H\left(\hspace{.3mm}\overline{\sS_{\ssA}}\hspace{.3mm}\right),
\end{eqnarray}
for all $\sA\in\as^{(\ell)}$ where $\overline{\sS_{\ssA}}
\stackrel{\rm def}{=}\sS-\sS_{\ssA}$. \QED
\end{de}

From (\ref{weak-ramp1.eq}) and (\ref{weak-ramp2.eq}) in Definition
\ref{partial-SS.def}, it holds that 
$H(\sS|\sA)=H(\sS_{\ssA}|\overline{\sS_{\ssA}}\sA)+
H(\overline{\sS_{\ssA}}|\sA)=H(\overline{\sS_{\ssA}})$, and hence, a PD
ramp SS scheme satisfies Definition \ref{ramp.def}.  

Note that a PD ramp SS scheme can be regarded as a SS scheme
with plural secrets. 
Conversely, if a SS scheme for plural secrets $\sS^{(L)}$ with access
structure $\AS^{(L)}$ is given, we can construct a corresponding access
structure of a PD ramp SS scheme for the secret
$\sS=\{S_1,S_2,\ldots,S_L\}=\{S^{(1)},S^{(2)},\ldots,S^{(L)}\}$ in the
following way: Assign each share set $\sA\subseteq\sV$ to the 
family $\as_\ell$ where $\ell$ is given by 
\begin{eqnarray}
\ell=\left|
\{\ell':\sA \in \as^{(\ell')}\in\AS^{(L)}\}
\right|.
\end{eqnarray}
Then, the tuple of families $\{\as_0,\as_1,\ldots,\as_L\}\DEF\AS_L$ can be
regarded as the access structure of the PD ramp SS scheme.

The difference between Definition \ref{plural-SS.def} and Definition
\ref{partial-SS.def} is summarized as follows: In Definition
\ref{plural-SS.def}, from a share set $\sA\subseteq\sV$, we can decrypt
a subset of secrets 
$\sS^{(L)}$, i.e., $\sS^{(\ssA)}$,  according to the access structure
$\AS^{(L)}$. However, in the PD ramp SS schemes defined in
Definition \ref{partial-SS.def}, a share set $\sA\in\as_\ell$ decrypts
some $\sS_{\ssA}$ which satisfies
(\ref{number_of_secrets}), i.e., $\sS_{\ssA}$ is not specified by the
access structure $\AS_L$.

We note that the amount of the leaked information about $\sS$  from a
share set $\sA\in\as_\ell$ is $(\ell/L) H(\sS)$ in PD ramp SS
schemes. Hence, in the sense of (\ref{1}), there is no
difference between Definition \ref{ramp.def} and Definition
\ref{partial-SS.def}. That is, both definitions guarantee the
same security in the case that $\sS$ is meaningless if some part of
$\sS$ is missing. However, if each part of $\sS$ has explicit
meaning, PD ramp SS schemes are not secure, and hence, not desirable. 

To overcome such defects,  Yamamoto defined 
strong ramp SS schemes as follows \cite{HY-ieice}\footnotemark\footnotetext{In
\cite{HY-ieice}, strong ramp SS schemes are defined for
$(k,L,n)$-threshold ramp access structures.}:

\begin{de}[\cite{HY-ieice}]\label{strong-SS.de}
Let $\sS=\{S_1,S_2,\ldots,S_L\}$ and $\AS_L$ be a secret and an access
 structure, respectively. Then, $\{\AS_L,\sV,\sS\}$ is called a {\em
 strong} ramp SS scheme if for all $\l=0,1,\ldots,L-1$, $\sA\in\as_\ell$
 satisfies (\ref{1}) and 
\begin{eqnarray}
\label{strong-ramp.eq}
H(S_{j_1}S_{j_2}\cdots S_{j_{L-\ell}}|\sA) 
= H(S_{j_1}S_{j_2}\cdots S_{j_{L-\ell}})
~\mbox{for all}~\{S_{j_1},S_{j_2},\ldots,S_{j_{L-\ell}}\}\subseteq \sS.
\end{eqnarray}
 \QED
\end{de}

Definition \ref{strong-SS.de} implies that strong ramp SS schemes do not
leak out any part of the secret explicitly from a non-qualified set
$\sA\not\in\as_L$. Now,
from this point of view, we review the $(k,L,n)$-threshold SS
scheme based on Shamir's interpolation method. 

\begin{rem}\label{shamir-insecure.rem}
We note that the $(k,L,n)$-threshold ramp SS scheme, which is an
 extension of Shamir's interpolation method
 \cite{S-cacm}, is not always a strong ramp SS scheme. For instance,
 consider a $(4,2,n)$-threshold ramp SS scheme by using the following
 polynomial of degree $3$ over the finite field $\mathbb{Z}_{17}$. 
\begin{eqnarray}
f(x)=S_1+S_2x+R_1x^2+R_2x^3,
\end{eqnarray}
where $\sS=\{S_1,S_2\}$ is a secret, and $R_1$ and $R_2$ are independent
 random numbers. The $i$-th share is given by 
 $V_i=f(i)$. Then, from a simple calculation of $V_3,V_6$ and
 $V_{15}$, we have
\begin{eqnarray}
5 S_2 = 7V_3 + 9V_6 +V_{15}.
\end{eqnarray}
This means that partial information $S_2$ can be decrypted completely
 from shares $V_3,V_6$ and $V_{15}$. 

We also note that from share set $\{V_1,V_2,V_3\}$, we have
 $H(S_\ell|V_1V_2V_3)=H(S_\ell)$ for $\ell=1,2$, and hence, the ramp SS
 scheme in this example is neither PD nor
 strong\footnotemark\footnotetext{In \cite{HKEY-trieice-03}, 
 a construction method is discussed for neither PD nor strong ramp SS
 schemes.}.\hfill $\Box$
\end{rem}

Remark \ref{shamir-insecure.rem} shows that it is difficult to construct
strong ramp SS schemes in general. In \cite{HY-ieice}, it is proposed
how to construct strong $(k,L,n)$-threshold ramp SS schemes, but it
is not known how to construct strong ramp SS schemes for general access
structures. 

Fortunately, PD ramp SS schemes with general access structure
$\AS_L$ can easily be constructed if $\AS_L$ satisfies monotonicity
given by (\ref{kurosawa.eq}) in Theorem \ref{KOSOT.thm}. Furthermore, it
is easy to calculate how much information leaks out from each
non-qualified set in PD ramp SS schemes. Therefore, we propose a method
to construct strong ramp SS schemes with general access structures based
on PD ramp SS schemes.

\section{Strong Ramp Secret Sharing Schemes 
with General Access Structures}
In this section, we propose how to construct a strong ramp SS scheme
with general access structure $\AS_L$ from a given PD ramp SS scheme
with the same access structure  $\AS_L$. 

Since a PD ramp SS scheme
with general access structure $\AS_L$ can always be constructed if
$\AS_L$ satisfies (\ref{kurosawa.eq}) in Theorem \ref{KOSOT.thm}, 
we assume that a PD ramp SS scheme with access structure
$\AS_L=\{\as_1,\as_2,\ldots,\as_L\}$ is obtained for a secret
$\sS=\{S_1,S_2,\ldots,S_L\}$. Denote
by $\phi_{\AS_L}(\sS,\sR)$ the encoder of such a PD ramp SS scheme with
the access structure $\AS_L$ for the secret $\sS$ where $\sR$ represents a set
of random numbers used in the encoder. Then, we choose publicly an $L
\times L$ non-singular matrix $T$ and define a new
encoder $\varphi_{\AS_L}(\sS',\sR)\DEF\phi_{\AS_L}(\sS'T,\sR)$ where
$\sS'=\{S'_1,S'_2,\ldots,S'_L\}$\footnotemark\footnotetext{
Hereafter, for simplicity of notation, we identify the sets 
$\sS=\{S_1,S_2,\ldots,S_L\}$ and $\sS'=\{S_1',S_2',\ldots,S_L'\}$ with
$L$-dimensional row vectors $[S_1~S_2 \cdots S_L]$ and
$[S'_1~S'_2\cdots S'_L]$, respectively.}.

The next theorem gives the necessary and sufficient condition of $T$
that realizes a strong ramp
SS scheme with the access structure $\AS_L$ for  secret
$\sS'=\{S'_1,S'_2,\ldots, S'_L\}$. 

\begin{thm}\label{main.thm}
Suppose that the encoder $\phi_\AS(\sS,\sR)$ of a PD ramp SS scheme with
an access structure $\AS_L$ for a secret $\sS$ is given. Let
 $\sS_{\ssA}$ be the partial information of the secret $\sS$ that can be
 decrypted explicitly 
 from a share set $\sA$ in the PD ramp SS scheme, and denote by
 $\sI(\sA)$ the set of indices of $\sS_{\ssA}$. 
Then, we construct a new encoder
 $\varphi_{\AS_L}(\sS',\sR)\DEF\phi_{\AS_L}(\sS'T,\sR)$ for a new secret 
$\sS'=\{S'_1,S'_2,\ldots,S'_L\}$ by using a publicly opened $L \times L$
 non-singular matrix $T$.

Then, the necessary and sufficient condition of $T$ to realize a strong ramp
 SS scheme $\{\sS',\sV,\AS_L\}$ is given by
\begin{eqnarray}
\label{condition.eq}
\mbox{\rm rank}~\left[T^{-1}\right]_{\hspace*{.45mm}\langle j_1,j_2,\ldots,j_{L-\ell} \rangle}
^{\langle \{1,2,\ldots,L\}-\mbox{\boldmath\scriptsize$I$}(\mbox{\boldmath\scriptsize$A$})\rangle}
=L-\l,
\end{eqnarray}
for all $\sA\in\as_\l$, $\l=0,1,\ldots,L$, where  
$\left[T^{-1}\right]^{\langle i_1,i_2,\ldots,i_u \rangle}_{\hspace*{.4mm}
\langle 
j_1,j_2,\ldots,j_u \rangle}$ is the submatrix that consists of the
 $i_1$-th, $i_2$-th$,\ldots,i_u$-th rows, and the $j_1$-th,
 $j_2$-th$,\ldots,j_u$-th columns of $T^{-1}$. \QED
\end{thm}

\begin{rem}
Theorem \ref{main.thm} implies that any strong ramp SS
schemes can be obtained from the corresponding PD ramp SS schemes
{\em without} loss of coding rates. \QED
\end{rem}

{\em  Proof of Theorem \ref{main.thm}:} Since the matrix $T$ is
non-singular, $\sS$ has one to one correspondence with $\sS'$. Hence,
$\sS'$ is also a set of $L$ mutually independent random variables according
to the same uniform distribution. Therefore, it holds that
$H(\sS)=H(\sS')=L\log|{\mathbb F}|$ where $\mathbb{F}$ is a finite
field in which $S_\ell$, $\ell=1,2,\ldots,L$ take values. 

Then, for any $\sA\in\as_\ell$, $\ell=1,2,\ldots,L$, where
$\AS_L=\{\as_0,\as_1,\ldots,\as_L\}$ is the access structure of the PD
ramp SS scheme, we have 
\begin{eqnarray}
H(\sS'|\sA)&=&H(\sS|\sA)
=\frac{L-\ell}{L}H(\sS)
=(L-\ell)\log|{\mathbb F}|=\frac{L-\ell}{L}H(\sS').
\end{eqnarray}
Therefore, (\ref{1}) holds for  secret $\sS'$. Next, from
(\ref{strong-ramp.eq}), we have for any 
$\{S'_{j_1},S'_{j_2},\ldots,S'_{j_{L-\ell}}\}\subseteq\sS'$ 
that 
\begin{eqnarray}
\nonumber
H(S'_{j_1}S'_{j_2} \cdots S'_{j_{L-\l}}|\sA)
 &=& H\left(\left.\sS\left[T^{-1}\right]_{\langle j_1,j_2,\ldots,j_{L-\l}
\rangle}^{\langle 1,2,\ldots,L\rangle}
\right|\sA\right) \\
\nonumber
 &\stackrel{\rm (a)}{=}&
 H\left(\left.\overline{\sS_{\mbox{\boldmath\scriptsize$A$}}}
\left[T^{-1}\right]_{\langle j_1,j_2,\ldots,j_{L-\l}\rangle}
^{\langle \{1,\ldots,L\}-\mbox{\boldmath\scriptsize$I$}
(\mbox{\boldmath\scriptsize$A$})\rangle}\right|\sA\right) \\
\nonumber
 &\stackrel{\rm (b)}{=}&
 H\left(\left.
\overline{\sS_{\mbox{\boldmath\scriptsize$A$}}}\right|\sA\right)\\
\label{23}
 &\stackrel{\rm (c)}{=}&
 H\left(\hspace*{.5mm}\overline{\sS_{\mbox{\boldmath\scriptsize$A$}}}
\hspace*{.5mm}\right)
= (L-\l) \log |{\mathbb F}|
=H(S'_{j_1}S'_{j_2} \cdots S'_{j_{L-\l}}), 
\end{eqnarray}
where equalities (a), (b), and (c) hold because of
(\ref{weak-ramp1.eq}), (\ref{condition.eq}) and (\ref{weak-ramp2.eq}),
respectively. 

Finally, we note that the necessity of (\ref{condition.eq}) is clear
since equality (b) in (\ref{23}) does not hold if (\ref{condition.eq})
is not satisfied. \QED

\halflineskip
From the proof of Theorem \ref{main.thm}, it is sufficient to choose the 
matrix $T$ satisfying, instead of the condition (\ref{condition.eq}),
that every submatrix of $T^{-1}$ has the full rank. We note that the
{\em Hilbert matrix} $T_H$ has such a property. Each element of an $L
\times L$ Hilbert matrix $T_H=[
t_{ij}]_{1 \le i \le L \atop  1 \le j \le L}$ is given by 
\begin{eqnarray}\label{Hilbert.eq}
 t_{ij}=\frac{1}{x_i+y_j},
\end{eqnarray}
where $x_i$ and $y_j$ must satisfy for all $i,j\in\{1,2,\ldots,L\}$ that  
\begin{eqnarray}\label{req1.eq}
x_i+y_j \neq 0.
\end{eqnarray}
Note that every submatrix of the Hilbert matrix is also a Hilbert
matrix, and the determinant of the matrix $T_H$ can be calculated as
follows:
\begin{eqnarray}
\det T_H = \frac
{\displaystyle
\prod_{1 \le i < j \le L}(x_i-x_j)
\prod_{1 \le i < j \le L}(y_i-y_j)
}
{\displaystyle
\prod_{i=1}^L \prod_{j=1}^L (x_i+y_j)
}.
\end{eqnarray}
Hence, it is clear that every submatrix of $T_H$ is non-singular if and
only if 
\begin{eqnarray}\label{req2.eq}
x_i \neq x_j~~\mbox{and}~~y_i \neq y_j
\end{eqnarray}
are satisfied for $i \neq j$ in addition to (\ref{req1.eq}). 
Since $|\mathbb{F}|$ is usually assumed to be sufficiently large in
ordinal ramp SS schemes, it is easy to choose $\{x_i\}_{i=1}^L$ and
$\{y_i\}_{i=1}^L$ satisfying (\ref{req1.eq}) and (\ref{req2.eq}). 

Then, from Theorems \ref{KOSOT.thm} and \ref{main.thm}, the following
theorem holds.

\begin{thm}\label{condition.thm}
A strong ramp SS scheme with access structure
 $\AS_L$ can be constructed if and only if each
 $\tilde{\as_\ell}$, $\ell=1,2,\ldots,L$, satisfies the monotonicity given
 by (\ref{kurosawa.eq}) of Theorem \ref{KOSOT.thm}. \QED 
\end{thm}

\begin{ex}
Note that matrices satisfying (\ref{condition.eq}) may exist besides
the inverse of Hilbert matrices. As an example, in the case of $L=2$ and
 $|{\mathbb F}| \ge 3$, we can use the following matrix $T^{\rm ex}$, the
 inverse of which is not a Hilbert matrix.
\begin{eqnarray}\label{sample.eq}
T^{\rm ex}=
\left[
\begin{array}{cc}
1 &  1 \\
1 & -1 \\
\end{array}
\right].
\end{eqnarray}
By using the matrix $T^{\rm ex}$ in (\ref{sample.eq}), the PD ramp SS scheme 
given by (\ref{ex.eq1})--(\ref{ex.eq3}) in Example \ref{Okada.ex} can be
 transformed into a strong ramp SS scheme with access structure
 $\AS^{\rm ex}_2$ given by (\ref{OkK1.eq1}) and (\ref{OkK1.eq3}) such that 
$
V_1=\{R_1,R_3\}, 
V_2=\{R_2,R_4\}, 
V_3=\{R_1+R_4+S'_1+S'_2,R_2+R_3+S'_1-S'_2\}$, and
 $V_4=\{R_1+S'_1+S'_2,R_2+S'_1+S'_2\}$.
It is easy to check that $\sV=\{V_1,V_2,V_3,V_4\}$ realizes a
 strong ramp SS scheme with access structure  $\AS^{\rm ex}_2$ for
 secret $\sS'=\{S'_1,S'_2\}$. 

We note here that, in the case of the access structure $\AS_2^{\rm ex}$
 in Example \ref{Okada.ex}, 
the minimum size of $\mathbb{F}$ is $2$  in order realize the PD ramp SS
 schemes for secret $\sS$ \cite{OK-acrypt94}, although
 $|\mathbb{F}|\ge 3$ is required to realize a strong ramp SS
 schemes for $\sS'$ if we use the transformation $T^{\rm ex}$ in
 (\ref{sample.eq}). In this way, the minimum size of $\mathbb{F}$ to
 realize strong ramp SS schemes generally becomes larger than that
 required to realize PD ramp SS schemes.  \QED
\end{ex}

\begin{rem}
Note that the matrix $T$ described in Theorem \ref{main.thm} is the 
 transformation from a PD ramp SS scheme to a corresponding strong ramp
 SS scheme. However, weak but not PD ramp SS schemes as shown in Remark
 \ref{shamir-insecure.rem} cannot always be transformed into strong
 ramp SS schemes by the matrix $T$ satisfying
 (\ref{condition.eq}). For example, consider the $(3,2,3)$-threshold
 ramp SS scheme given by $V_1 = S_1+R,V_2 = S_1+S_2+R$, and $V_3 = R$,
where $R$ is a random number  \cite{HY-ieice}. Then, these shares
 realize a weak but not PD ramp SS scheme. If we transform this ramp SS
 scheme by using $\sS=\sS'T^{\rm ex}$ where $T^{\rm ex}$ is given by
 (\ref{sample.eq}), we have $V_1 = S'_1+S'_2+R, V_2 = 2 S'_1 + R$, and
 $V_3 = R$. It is easy to check that $V_1,V_2$ and $V_3$
 do not realize a strong ramp SS scheme for $\sS'$. 
\QED
\end{rem}

\end{document}